\documentclass{acmsiggraph}                     

\usepackage{booktabs} 
\usepackage{microtype}
\usepackage{amsmath}
\usepackage{amssymb}
\usepackage{units}
\DeclareMathAlphabet{\mathcal}{OMS}{cmsy}{m}{n}  
\usepackage{xfrac}
\usepackage[normalem]{ulem}
\usepackage{graphicx}  		
\usepackage{float}
\usepackage{url}
\usepackage{xspace}
\usepackage{color}
\usepackage[normalem]{ulem}  
\usepackage{enumitem}  		
\usepackage{tikz}			
\usepackage{calc}			
\usepackage{collcell}
\usepackage{tabularx}

\usepackage{multirow}
\usepackage{amsmath}
\usepackage{amssymb}
\usepackage{amsfonts}
\usepackage{eucal}
\usepackage[latin1]{inputenc}
\usepackage[normalem]{ulem}
\usepackage{listings}
\usepackage{mathrsfs}
\usepackage{subcaption}
\usepackage{cancel}
\usepackage[normalem]{ulem}
\usepackage{cancel}
\usepackage{tcolorbox}
\usepackage{verbatimbox}

\usepackage{graphicx}
\usepackage{color}
\usepackage{wrapfig}
\usepackage{ifthen}
\usepackage[]{algorithm2e}

\title{Importance Sampling of Many Lights with Reinforcement Lightcuts Learning}

\author{Jacopo Pantaleoni\thanks{e-mail: jpantaleoni@nvidia.com}\\NVIDIA}

\keywords{global illumination, light transport simulation, Monte Carlo, reinforcement learning}

\begin{document}
	
	\newcommand{\figdir}{figures}

	\teaser{
		\includegraphics[width=88.0mm]{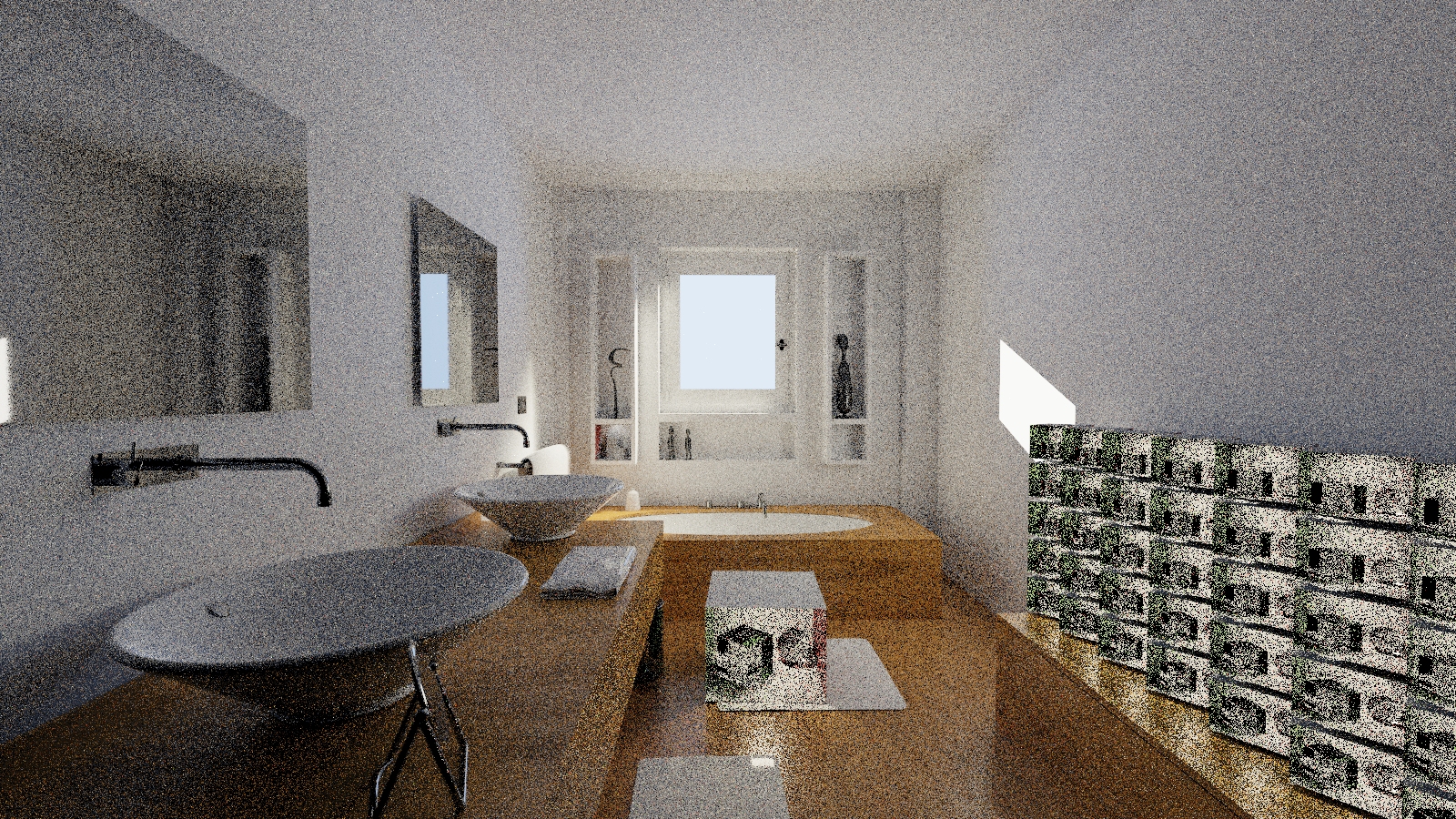}
		\includegraphics[width=88.0mm]{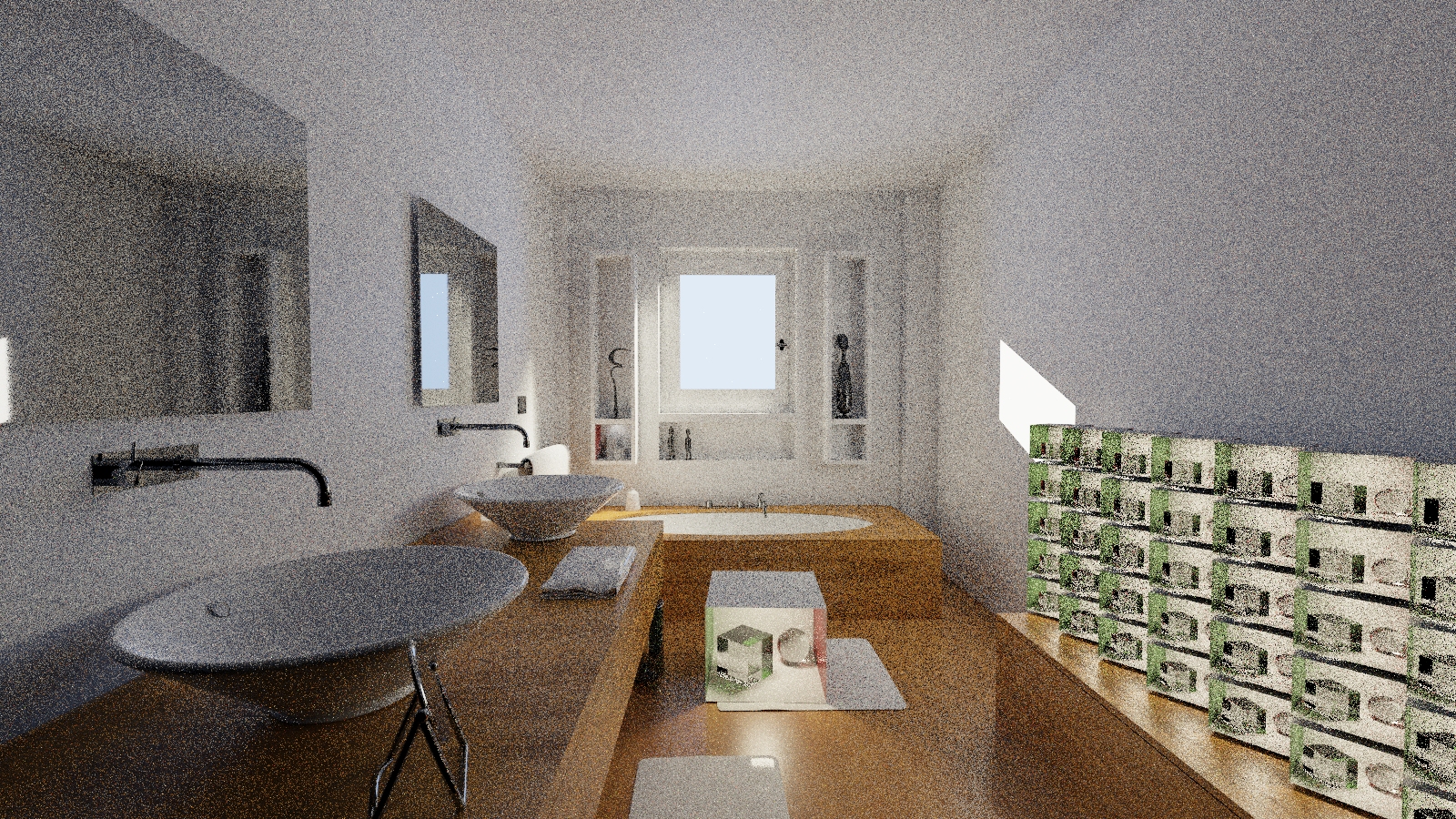}
		\caption{
			\emph{Cornell Boxes in Salle de Bain:}
			A scene featuring complex occlusion (each Cornell Box contains an area light that contributes to its own illumination, but not that of the neighboring ones), rendered with Path-Space Filtering with standard mesh light sampling (left) and with Reinforcement Lightcuts Learning (right), at 64spp.
		}
	}
	
	
	\maketitle
	
	
	\begin{abstract}
		
		In this manuscript, we introduce a novel technique for sampling and integrating direct illumination in the presence of many lights. Unlike previous work, the presented technique importance samples the product distribution of radiance and visibility while using bounded memory footprint and very low sampling overhead. This is achieved by learning a compact approximation of the target distributions over both space and time, allowing to reuse and adapt the learnt distributions both spatially, within a frame, and temporally, across multiple frames. Finally, the technique is amenable to massive parallelization on GPUs and suitable for both offline and real-time rendering.
		
	\end{abstract}
	
	
	\begin{CRcatlist}
		\CRcat{I.3.2}{Graphics Systems C.2.1, C.2.4, C.3)}{Stand-alone systems};
		\CRcat{I.3.7}{Three-Dimensional Graphics and Realism}{Color,shading,shadowing, and texture}{Raytracing};
	\end{CRcatlist}

	\keywordlist
	
	
	\copyrightspace

\section{Introduction}

Direct lighting simulation is a critical component of modern rendering algorithms, and while by construction it is simpler than solving the full rendering equation, it can easily be a challenging problem by itself. This is especially true in the context of the so called \emph{many lights} problem, where the scene is illuminated either by many distinct lights, or by mesh emitters, where the triangles in the mesh act as a complex collection of emitters. While in recent years several algorithms have been proposed to tackle these situations and improve the sampling efficiency by using various forms of importance sampling, the majority of them importance sample only the unoccluded contribution, hence suffering from unbounded variance in the case of complex occlusion, while the few approaches which also incorporate visibility suffer from very large and often unbounded memory requirements, directly proportional to the number of lights in the scene. Moreover, most of the approaches have historically targeted offline film rendering, and generally suffer from high per-sample computing overheads, often requiring one or more expensive memory-divergent tree traversals for the generation of each individual sample.
In this work we try to address all of these problems by presenting new algorithms that try to learn compressed importance sampling distributions approximating the fully occluded expected contributions of all light sources at arbitrary points in the scene. As the proposed representations utilize a bounded memory footprint, they serve the dual purpose of being both memory efficient and cheap to sample from, as they require a low constant number of memory accesses.
Furthermore, as the learning process adapts the learnt distributions over time, the algorithm is effective even for real-time rendering of dynamic lighting scenarios.

\begin{figure}
	\fbox{
		\includegraphics[width=41.0mm]{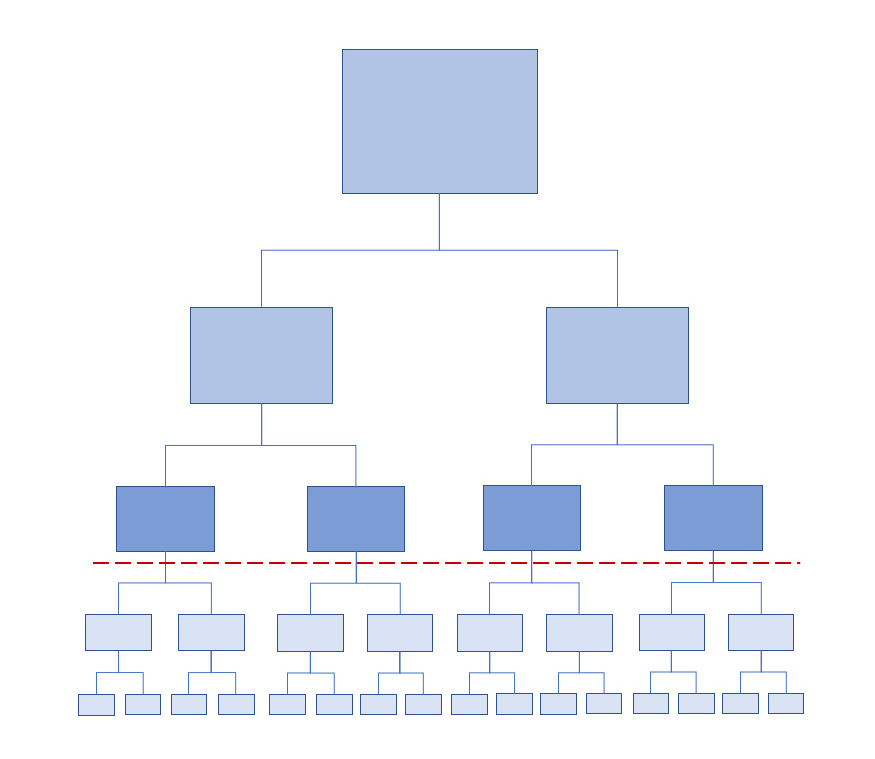}
		\includegraphics[width=41.0mm]{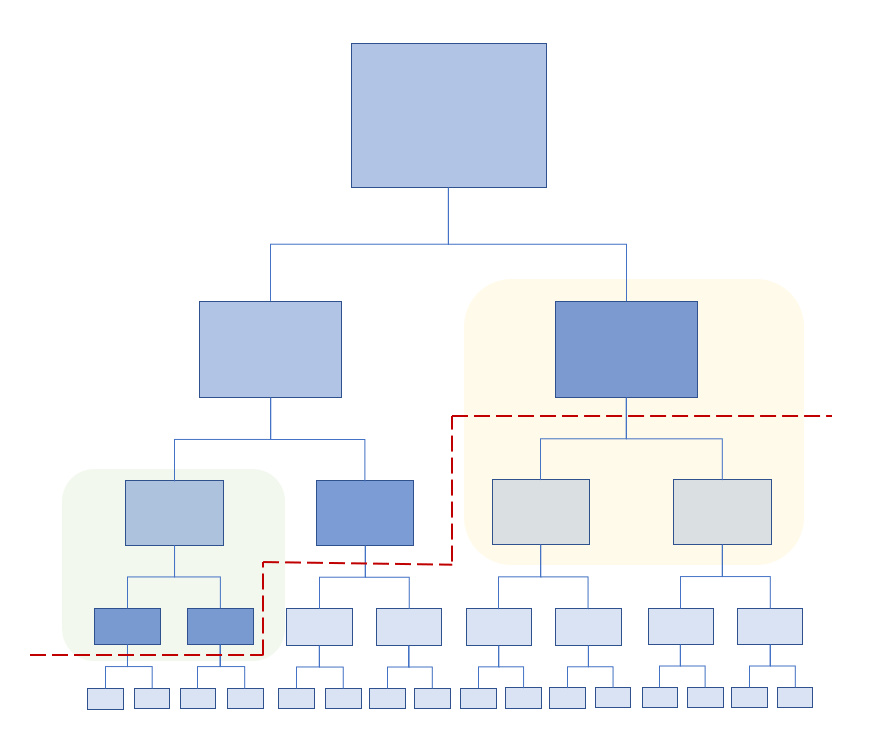}
	}
	\caption{Tree cuts, representing compact light clusterings, are adapted over time using our \emph{split-collapse} algorithm: the pictures show a balanced tree cut containing four leaves on the left, which is adapted on the right by splitting one leaf (highlighted in green), and collapsing one parent (highlighted in yellow). }
	\label{TreeCuts}
\end{figure}

\section{Previous Work}

One of the first techniques to address the many lights problem is Lightcuts, from Walter et al \shortcite{Walter:2005:LSA}. Their algorithm introduced the idea of using a global hierarchy over all lights, and using that to drive a local clustering of the lights for each shading point, by selecting a suitable \emph{cut} into the tree.
In order to achieve this, the algorithm computes bounds on the maximum potential contribution of each cluster to the shading point, and keeps splitting the most influential clusters until a user-defined error threshold is reached, in a process similar to that employed by \emph{tree codes} algorithms like that proposed by Barnes and Hut \shortcite{1986Natur} for N-body simulation.

While Lightcuts is targeting the computation of fixed error-bounds estimates, extending the algorithm to construct unbiased sampling estimators is relatively straightforward, as it has been independently shown by Conty and Kulla \shortcite{Conty:2018:ISM}, Vevoda and Krivanek \shortcite{Vevoda:2016:ADI} and more recently by Yuksel \shortcite{Yuksel:2019:SLC} and Moreau et al \shortcite{Moreau:2019:DML}.
Unfortunately, neither the original Lightcuts nor these importance sampling adaptations incorporate visibility in the calculation of the importance bounds assigned to tree nodes, essentially basing all sampling decisions on their expected \emph{unoccluded} contribution. In scenes featuring complex visibility, this shows as unbounded complexity in the original error-bound algorithm, and as unbounded variance in its importance sampling adaptations.

To our knowledge, the first algorithm that used visibility to drive importance sampling of light sources based on their expected occluded contribution is the Importance Caching work by Georgiev et al \shortcite{Georgiev:2012:IC}. Their approach is to cache one or more full cdfs of all lights in the scene at sparse locations, representing both the unoccluded and the occluded contributions of each individual light at the selected cache location.
At other nearby locations, samples are drawn proportionatley to the nearby cached cdfs, using multiple importance sampling \cite{Veach:PHD} to combine them.
While very effective, this algorithm is not very scalable due to the unbounded storage requirements (for each cache location, $O(N)$ in the number of lights $N$).

Another more recent approach that incorporates visibility is the reinforcement learning algorithm proposed by Dahm and Keller \shortcite{Dahm:2017}. In their algorithm, the cached cdfs are learnt over time using a variation of \emph{Q}-learning. This approach has the advantage of learning the actual expected contribution \emph{across the entire cache cell}, as opposed to using the contribution at a single point within the cell. Moreover, as the learning process is distributed across all shading points, the algorithm is computationally more scalable, although the memory scalability concerns remain the same, as the Q-tables are still O(N) in the number of lights.

Our algorithm draws many similarities to all these approaches.
Similarly to Vevoda and Krivanek \shortcite{Vevoda:2016:ADI} we use a grid structure to cache lazily computed tree cuts over a light hierarchy; however, unlike theirs, our cuts are bounded in size, and most importantly the importance estimates assigned to the nodes in the cut are learnt using their actual, occluded contribution, employing a variant of the reinforcement learning algorithm suggested by Dahm and Keller \shortcite{Dahm:2017}. Moreover, our algorithm adapts the selected cuts over time, essentially learning to use the most appropriate light clustering for each cell.

\section{Main contribution}

Our algorithm merges four fundamental ingredients.
The first is a 5d hash-table which represents a sparse 5d grid on the space of shading points, where the first 3 dimensions encode for positions and the last 2 encode for normals. The algorithm we use in order to hash shading points is loosely based on the jittered spatial hashing described by Binder et al \shortcite{Binder:2018:FPS}, where the hash cells are located by jittering the quantized position and normal components of the shading point. While the normals are quantized using fixed precision, in our path tracer the positions are quantized according to cell sizes derived from ray footprints calculated based on the area pdf of the shading points (i.e. the path vertices) themselves. The latter allows to use cells that are roughly as large as a user defined tile for visible points, and become coarser for subsequent bounces.

The second key ingredient is the compact representation we employ to store importance caching distributions at each hashed grid cell. While previous algorithms opted either for a hierarchy on all lights used to implicitly encode the probability of each node, computed on the fly without taking visibility into account \cite{Conty:2018:ISM,Vevoda:2016:ADI}, or for explicit tables storing the full cdf of the shadowed contribution of all lights at some representative shading point within each cell \cite{Georgiev:2012:IC}, we use a hybrid data structure with bounded memory footprint. The idea is to have each cell encode a fixed-size clustering of all lights, and use the clustering to store a compact cdf across the clusters only - in practice replacing the dense cdf by a compressed representation.
A distinguishing trait of our approach is that both the clustering and the cdf assigned to the selected set of clusters are learnt and adapted over time independently at each cell location.

In order to achieve this, we constrain all the possible clusterings to a set that allows for a particularly simple representation: fixed size \emph{tree cuts}. In practice, we sort all lights in the scene and build a single binary tree over all of them in a preprocess, and use that \emph{reference} tree to guide the selection and representation of a potentially different fixed size cut for each hash cell.
The leaves of each cut will represent a partitioning in disjoint light clusters.
Notice that due to the tree representation used as an \emph{implicit guide} each cut can be represented either as a list of node indices or, since the lights are sorted in tree order, as a list of contiguous ranges into the sorted lights array. Particularly, since the ranges are contiguous we can represent each cut as a monotonically increasing list of indices marking the end of each cluster.
The initial cut used when a cell is first created can be chosen by any arbitrary heuristic, for example descending the tree in breadth first order until the desired amount of nodes $M$ is collected.
In practice, we build such a cut once and simply copy it when a cell is first accessed.

The third key ingredient is the learning procedure used to assign probabilities to the leaves of each cut. The technique we use to do this is an adaptation of the reinforcement learning approach described by Dahm and Keller \shortcite{Dahm:2017}. Together with the cut in each hash cell, we store two separate tables $Q$ and $C$ representing respectively the \emph{Q}-learning table over its leaves (essentially encoding a pdf) and a derived cdf.
While rendering a frame (or a rendering pass, if a frame is split in sub-frames), at each shading point, whenever we need to sample direct illumination, we use the cdf in the corresponding cell to sample a light cluster $s$, and then select a light in the cluster randomly, $l \in s$.
Finally, we compute the contribution $c_l$ of the selected light source and use it to update the $Q$-table accordingly with the reinforcement learning rule:
\begin{equation}
Q(s) := (1 - \alpha) \cdot Q(s) + \alpha \cdot |c_l|
\end{equation}
The cdf $C$ of each hash cell is derived from the corresponding $Q$-table at the end of the rendering pass, according to the equation:
\begin{eqnarray}
C(0) &=& Q(0) \nonumber \\ 
C(s) &=& C(s-1) + Q(s).
\end{eqnarray}

Finally, the last key ingredient of our approach is the algorithm we use to adapt the cuts across rendering passes.
Since we potentially need to update tens of thousands such cuts per frame, and we want our approach to be usable in real-time scenarios, we devised a very simple and highly scalable algorithm.
The idea is to look at the probabilities assigned to the leaves $L$ in each cut, and at those of their parents $P$, and determine which leaves to split and which to merge based on those. In practice, we use the following heuristic: we consider the leaf with highest probability, $l_{max} = argmax_{l \in L}(Q(l))$ and the parent with lowest probability, $p_{min} = argmin_{p \in P}(Q(p))$; if $Q(l_{max}) > T \cdot Q(p_{min})$, where $T$ is a small constant, we split the leaf $l_{max}$ into its two children, and collapse the leaves of the interior node $p_{min}$, replacing them by $p_{min}$ itself.
We call this basic algorithm \emph{split-collapse}, and observe that, at least in theory, we could also invoke it iteratively until no pairs of nodes are touched anymore. This would roughly correspond to making the leaf weights as similar as possible. The key step is shown schematically in Figure~\ref{TreeCuts}.

\subsection{Implementation Details}

We implemented the above algorithms in Fermat using CUDA C++, and tested them on an NVIDIA RTX 2080 Ti.
For the global tree creation, we use a simple LBVH builder using the 3d positions of the lights only. A more advanced hierarchy construction method considering the light source orientations and perhaps a surface area heuristic similar to that described by Conty and Kulla \shortcite{Conty:2018:ISM} could be used instead, though we left this as future work.

The cdf computation described in equation 2 is realized through a parallel prefix-sum, launching a single cooperative thread array, or thread group, per cell/cut.

Similarly, we implemented the \emph{split-collapse} algorithm as another parallel kernel, again launching a cooperative thread array per cell/cut.
The kernel involves three broad phases: the first to compute the pdf of each internal node in the cut, starting with one thread per leaf and looping through the parents till the root is reached, and using shared memory atomics to update the corresponding values; the second employing two parallel prefix-sums, one to compute the maximum probability among the leaves and another to compute the minimum among the parents; and finally the third to compare the two resulting values and eventually emit the output cut with the split and merged leaves replaced.

At render time, sampling the cdf is performed by a standard binary search. Since the cdf is bounded in size, this typically involves a few steps only.
In all our experiments we employed either 128 or 256 leaves per cut.

\section{Results}

\begin{figure*}
	\includegraphics[width=88.0mm]{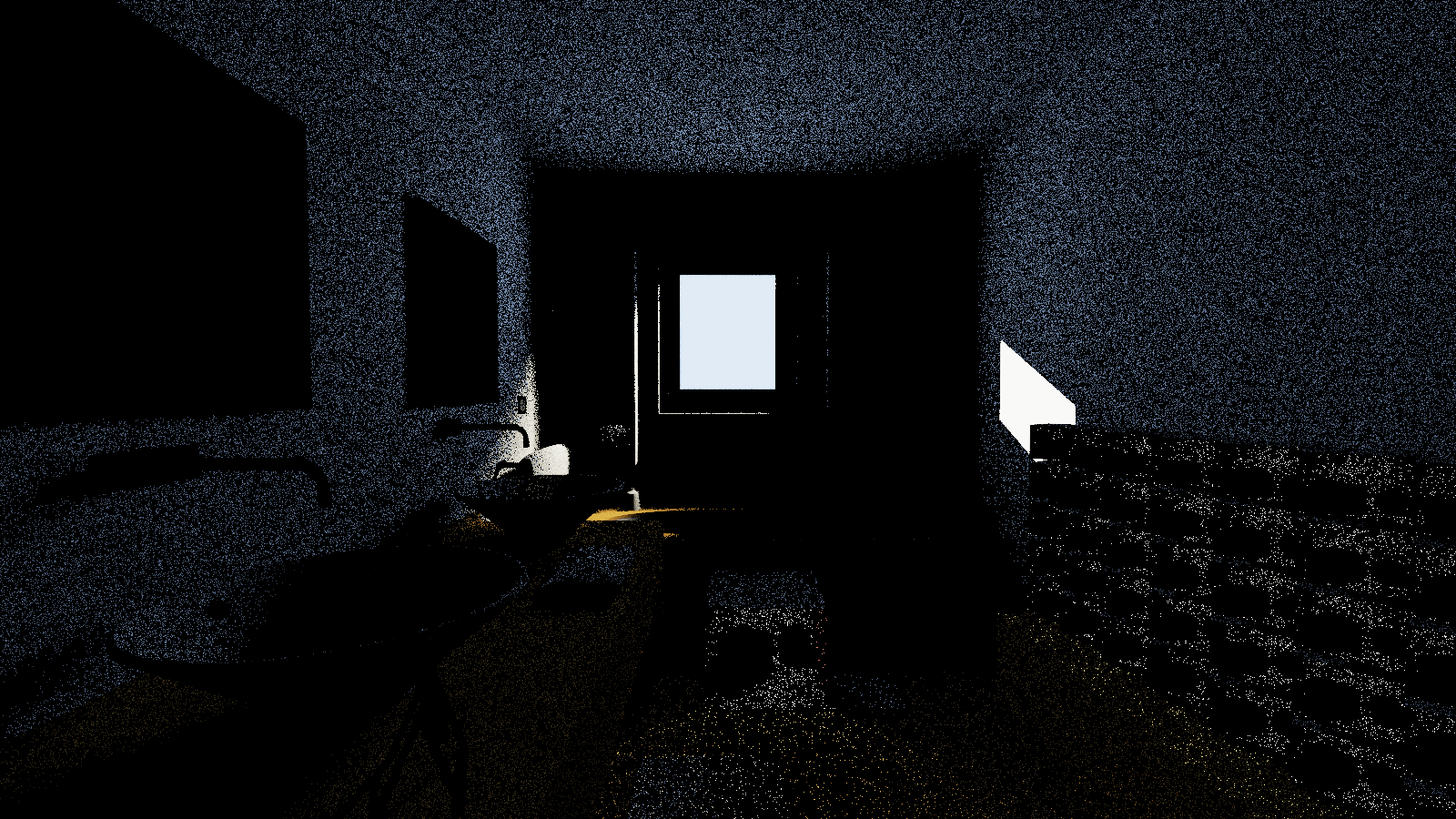}
	\includegraphics[width=88.0mm]{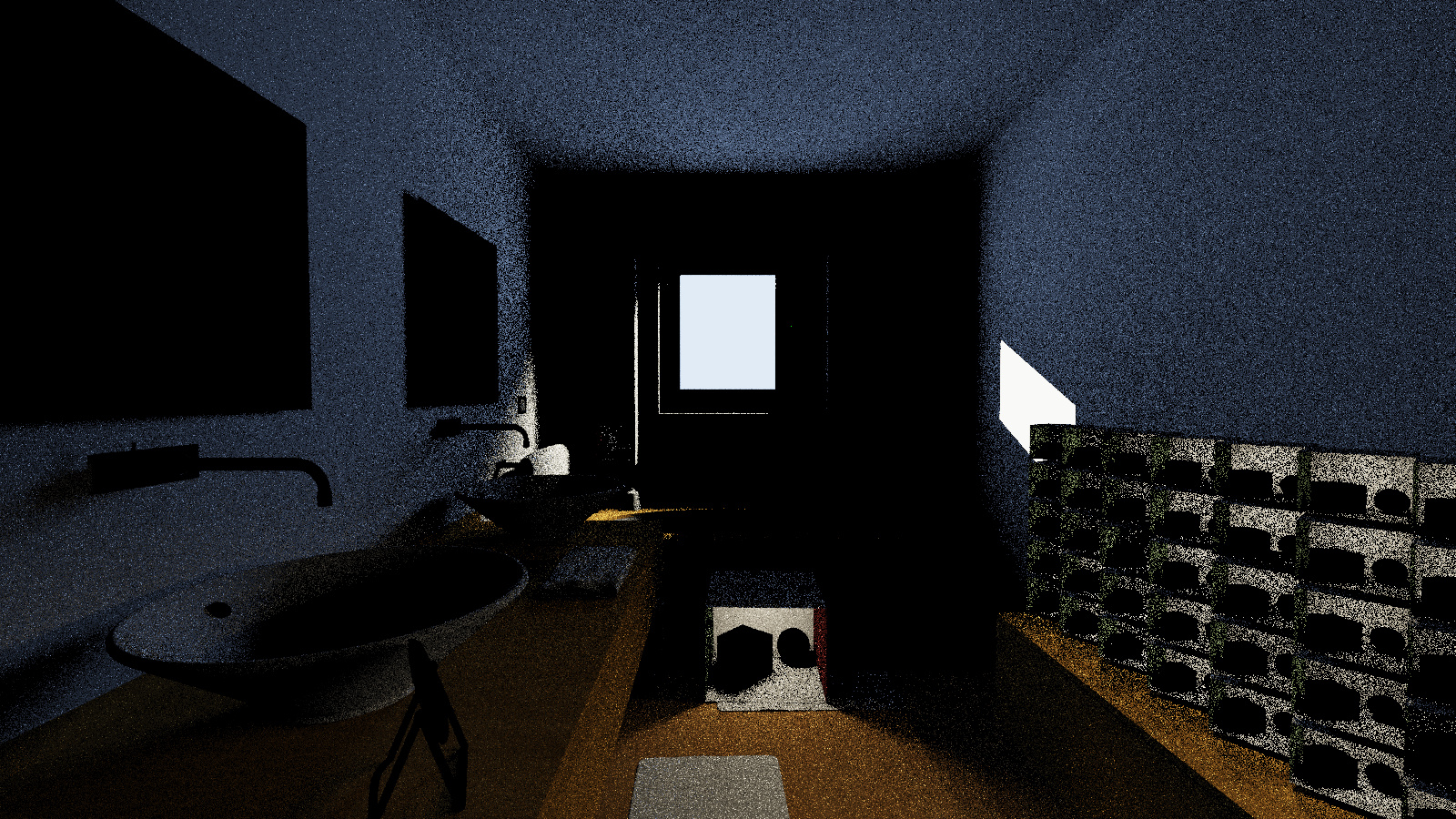} \\
	\includegraphics[width=88.0mm]{results/pt-mesh-128spp}
	\includegraphics[width=88.0mm]{results/pt-rl-128spp}
	\caption{Cornell Boxes in Salle de Bain, renderered with direct lighting only, at 8spp (first row),
		and simple path tracing, at 128spp (second row).
		Left column: standard mesh light sampling. Right column: Reinforcement Lightcuts Learning. }
	\label{SalleDeBain}
\end{figure*}

We compared our technique against an energy-based importance sampler, which samples triangles proportionately to their total emission (i.e. the integral of emission intensity over the area) using cdf inversion.
Figure~\ref{SalleDeBain} shows a comparison on a scene featuring both sky light illumination from an environment dome, triangulated into 2048 triangles, as well as illumination from 42 local area lights modeled as 2 triangles each, each illuminating a small Cornell Box.
Importance sampling methods which ignore visibility \cite{Conty:2018:ISM,Yuksel:2019:SLC,Moreau:2019:DML} would all fail in this scene, as each local light illuminates its own Cornell Box only and not the others.
Our technique is able to vastly reduce sample variance and accelerate convergence due to its capabilities to quickly learn global visibility in an adaptive fashion.

\begin{figure*}
	\includegraphics[width=88.0mm]{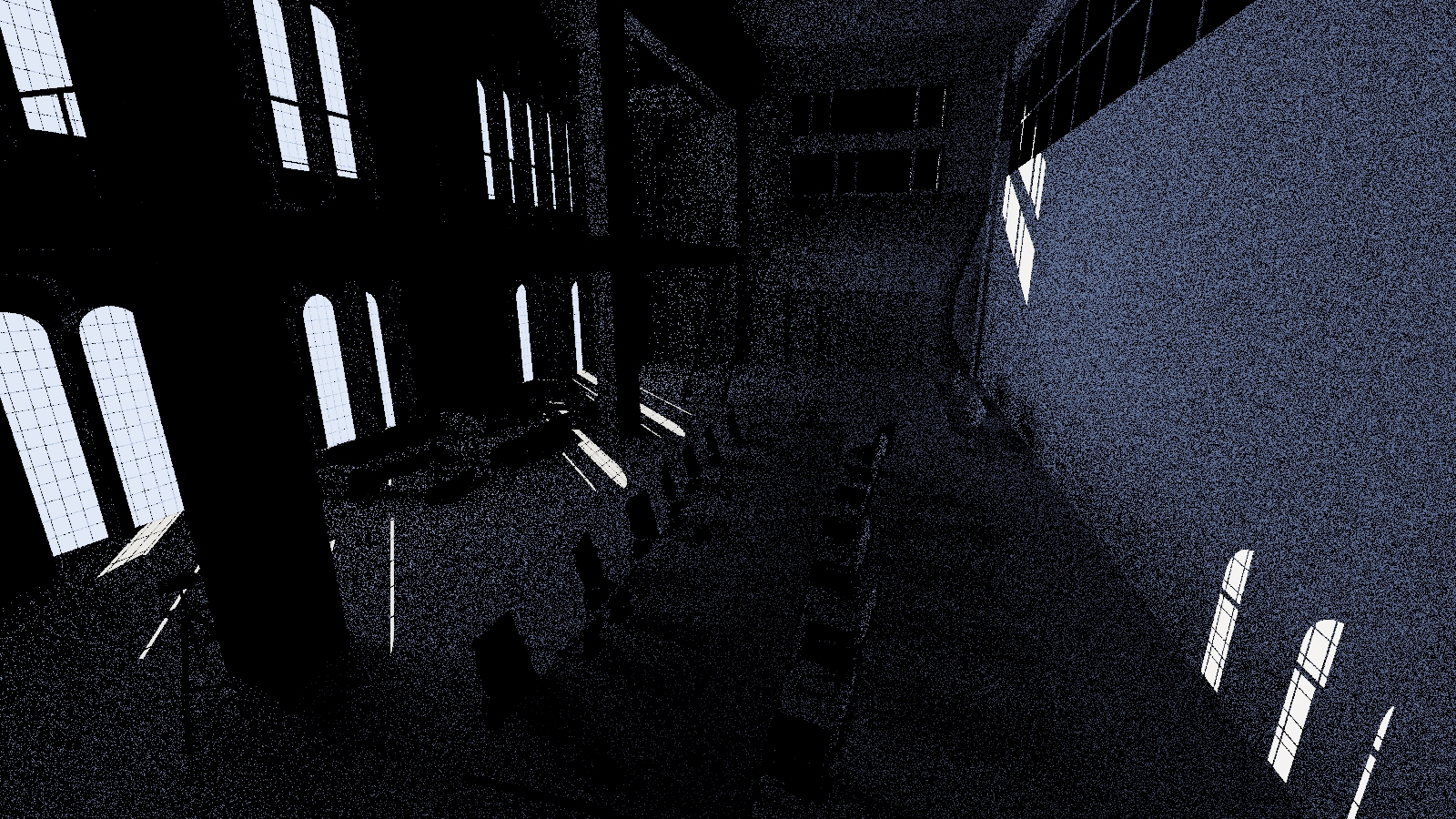}
	\includegraphics[width=88.0mm]{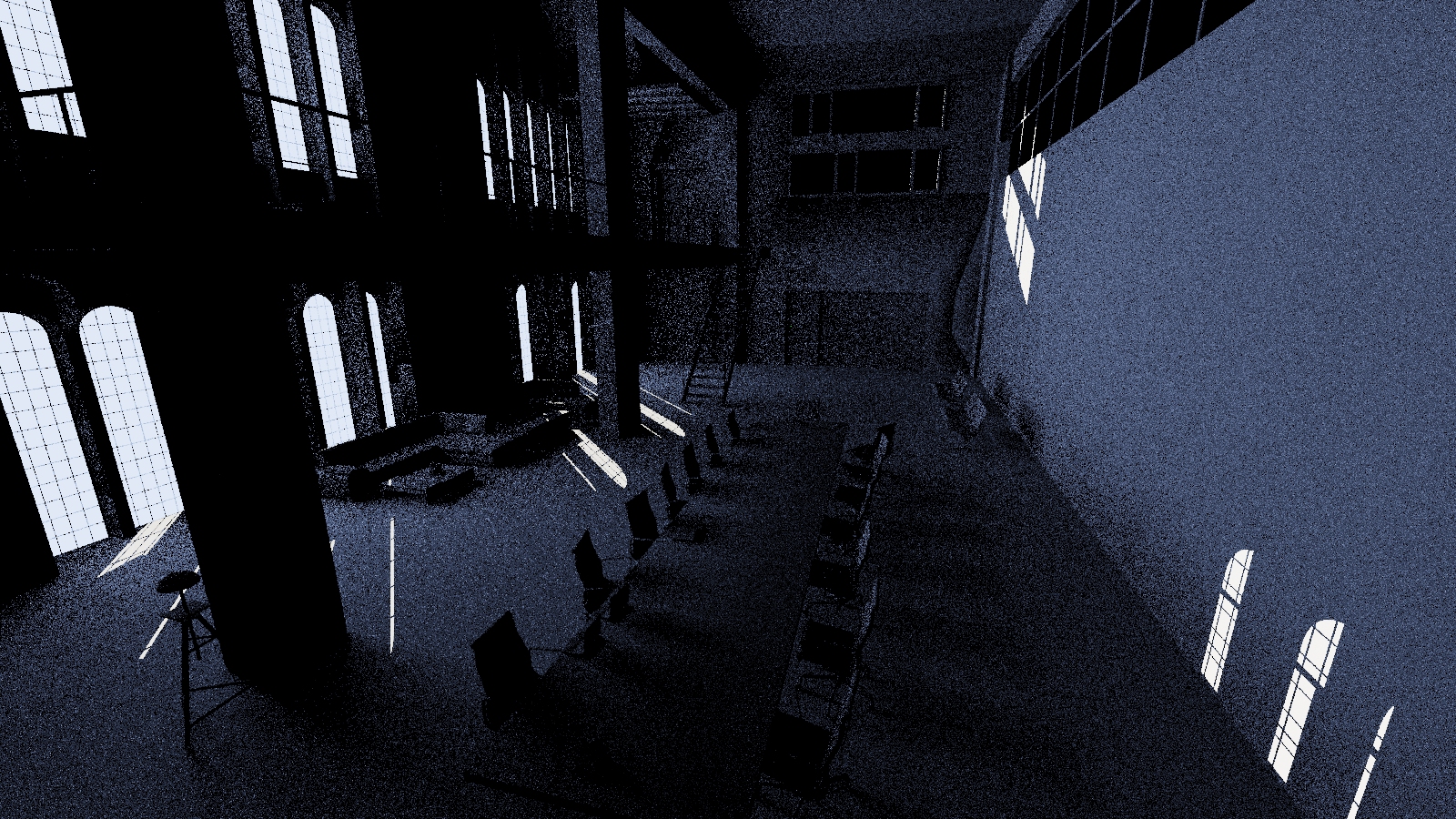} \\
	\includegraphics[width=88.0mm]{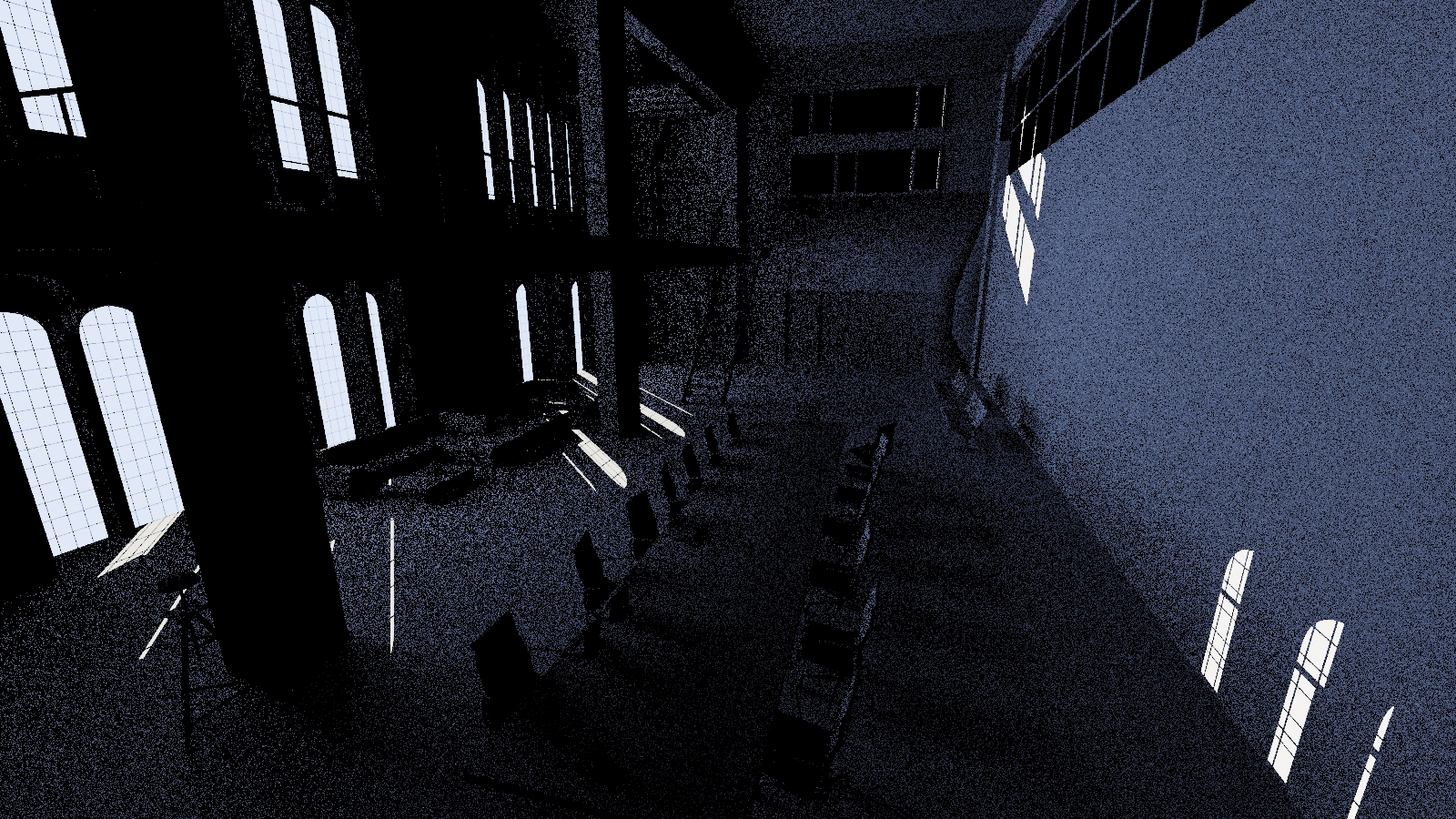}
	\includegraphics[width=88.0mm]{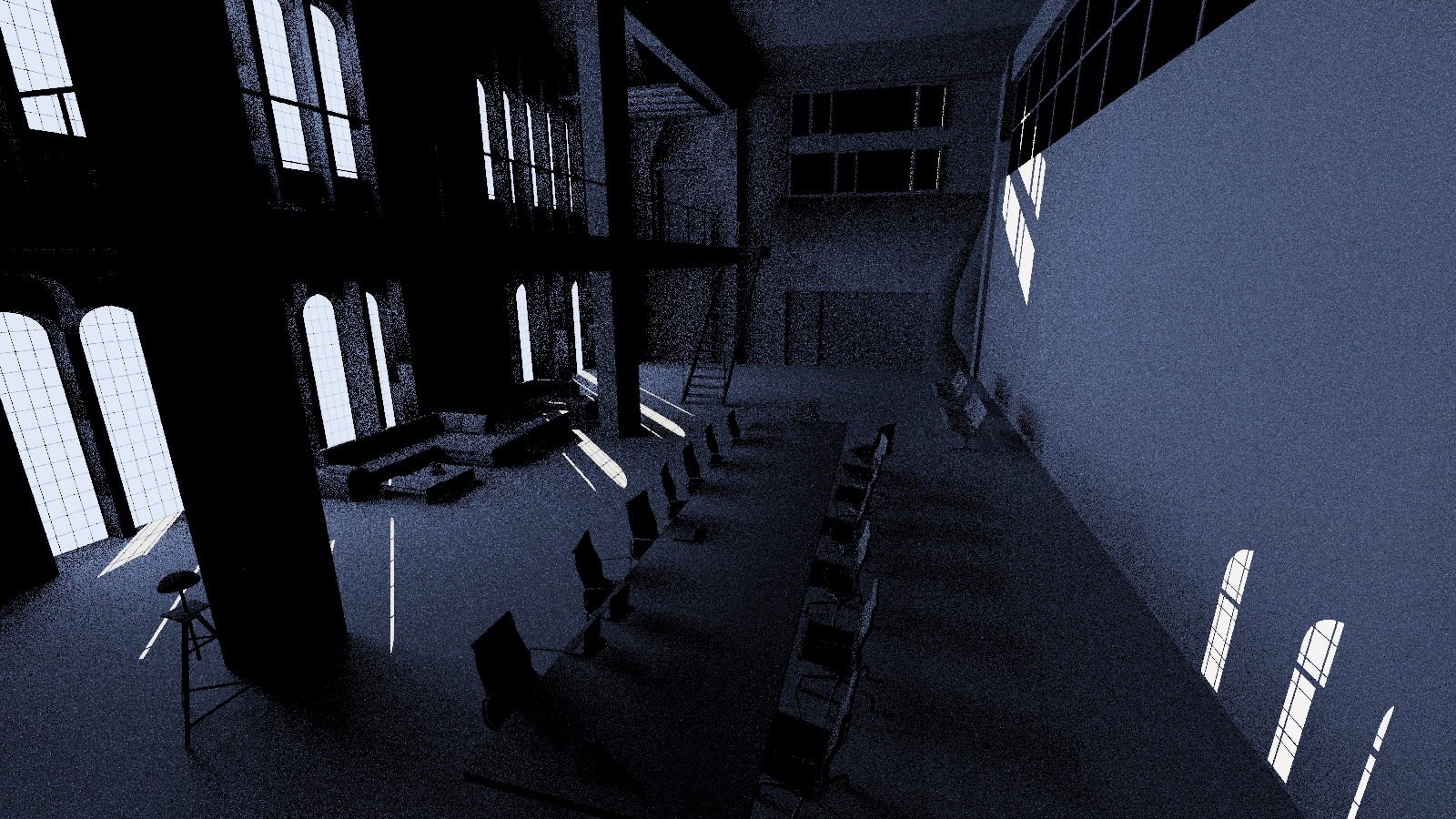}
	\caption{Complex visibility at play: this room is illuminated by the sky and a directional light only. Most of the sky is shadowed, and only a part of it is visible through the window openings. Upper row: 4spp. Bottom row: 8spp.\\
	Left column: standard mesh light sampling. Right column: Reinforcement Lightcuts Learning }
\label{ArchInterior}
\end{figure*}

Figure~\ref{ArchInterior} shows another scene in which a uniform sky light is the only source of illumination, but where again visibility plays a crucial role due to the presence of many small windows.
Since the emissive geometry is at approximately equal distance from all visibile points, in this case any importance sampling method based solely on the BRDF and unoccluded radiance is theoretically bound to fail, and not improve variance compared to a purely uniform sampler. Our approach is again able to reduce variance by roughly an order of magnitude.
In both cases our implementation has an overhead of about 4 to 12ms for full path-tracing at 1080p, depending on the settings.

\section{Discussion}

We have presented a simple algorithm that is able to dynamically learn to importance sample from large collections of light sources for next-event estimation. The algorithm has modest overhead and is massively scalable, hence being well-suited for real-time ray-tracing applications.

Unlike most previous many-lights algorithms targeting real-time, our algorithm is importance sampling the true target occluded distribution, incorporating the often crucial visibility term.

Its core component is a novel massively parallel algorithm for adaptive refinement of succint \emph{tree cuts} into a reference tree, allowing to express and adapt complex distributions with a very low memory footprint.
We believe this novel representation will be useful for many other rendering applications.

\bibliographystyle{acmsiggraph}
\bibliography{main}

\end{document}